\documentclass[12pt]{article}

\def\bl{\begin{eqnarray}}
\def\el{\end{eqnarray}}
\def\bll{\begin{eqnarray*}}
\def\ell{\end{eqnarray*}}
\usepackage{amssymb}
\usepackage{amsmath,amsthm}
\usepackage{amsfonts}
\usepackage{fleqn}
\usepackage{titlesec}
\usepackage[backref=true,dvipdfm,draft=false,colorlinks=true,
bookmarksopen=true,bookmarks=true,bookmarksnumbered=false,
linkcolor=blue,citecolor=blue,urlcolor=blue,pdfauthor=YunBaoHuang]{hyperref}

\title{Smooth infinite words over $n$-letter alphabets having same remainder when divided by $n$}

\small \author{Yun Bao Huang \\
 Department of Mathematics\\
 Hangzhou Normal University\\
 Xiasha Economic Development Area\\
 Hangzhou, Zhejiang 310036, China\\
        huangyunbao@sina.com\\
        huangyunbao@gmail.com}
\date{2010.11.18}
\begin{document}
\numberwithin{equation}{section}
\makeatletter
\titlelabel{\thetitle.\,}
\newcommand{\Extend}[5]{\ext@arrow 0099{\arrowfill@#1#2#3}{#4}{#5}}
\newcommand{\prf}[1]{\noindent\bf{Proof.}\;\;\rm{#1}}
\newtheorem{defn}{Definition}
\newtheorem{thm}[defn]{Theorem}
\newtheorem{lem}[defn]{Lemma}
\newtheorem{prop}[defn]{Proposition}
\newtheorem{cor}[defn]{Corollary}
\newtheorem{conj}[defn]{Conjecture}
\newtheorem{exmp}[defn]{Example}
\newtheorem{rem}[defn]{Remark}
\makeatother \maketitle
\begin{quote}
{\small {\bf Abstract.} Brlek et al. (2008) studied smooth infinite
words and established some results on letter frequency, recurrence,
reversal and complementation for 2-letter alphabets having same
parity. In this paper, we explore smooth infinite words over
$n$-letter alphabet $\{a_1,a_2,\cdots,a_n\}$, where $a_1<a_2<\cdots
<a_n$ are positive integers and have same remainder when divided by
$n$. And let $a_i=n\cdot q_i+r,\;q_i\in N$ for $i=1,2,\cdots,n$,
where $r=0,1,2,\cdots,n-1$. We use distinct methods to prove that
(1) if $r=0$, the letters frequency of two times differentiable
well-proportioned infinite words is $1/n$, which suggests that the
letter frequency of the generalized Kolakoski sequences is $1/2$ for
2-letter even alphabets; (2) the smooth infinite words are
recurrent; (3) if $r=0$ or $r>0 \text{ and }n$ is an even number,
the generalized Kolakoski words are uniformly recurrent for the
alphabet $\Sigma_n$ with the cyclic order; (4) the factor set of
three times differentiable infinite words is not closed under any
nonidentical permutation. Brlek et al.'s results are only the
special cases of our corresponding results.

}
\end{quote}
\newpage
\section{Introduction\label{s1}}
The Kolakoski sequence $K$ which was first introduced
in~\cite{Kolakoski}, is the infinite sequence over the alphabet
$\Sigma=\{1,2\}$ which starts with 1 and equals the sequence defined
by its run lengths:

$K=
\underbrace{1}_{1}\underbrace{22}_{2}\underbrace{11}_{2}\underbrace{2}_{1}
 \underbrace{1}_{1}\underbrace{22}_{2}\underbrace{1}_{1} \underbrace{22}_{2}\underbrace{11}_{2}
 \underbrace{2}_{1}\underbrace{11}_{2}\underbrace{22}_{2}\underbrace{\cdots}_{\cdots}$\\
Here, a run is a maximal subsequence of consecutive identical
symbols. The curious Kolakoski sequence $K$ has received a striking
attention by showing some intriguing combinatorical properties
\cite{Br2,Br5,Carpi1,Carpi2,Dekking1,Huang1,Huang2,Huang3,
Lepist,Paun,Steacy,Steinsky,Weakly}.

     Keane~\cite{Keane} asked whether the density of $1'$s
in $K$ is 0.5. Chv\'{a}tal~\cite{Chv} proved that the upper density
of $1'$s as well as the upper density of $2'$s in $K$ is less than
0.501. Brlek, Jamet and Paquin~\cite{Br3} investigated smooth
infinite words on 2-letter alphabets having same parity and showed
that all smooth infinite words are recurrent; that the closure of
the set of factors under reversal holds for odd alphabets only; and
that the frequency (density) of letters in extremal words is 1/2 for
even alphabets, and for $a = 1$ with $b$ odd, the frequency of $b$'s
is $1/(\sqrt{2b-1}+ 1)$.

     Baake and Sing~\cite{Ba1} and Sing~\cite{Sing2, Sing3} established a
connection between the generalized Kolakoski words and model sets.
By the way, Sing~\cite{Sing2, Sing3} showed that for 2-letter
alphabets having same parity, the generalized kolakoski sequences
are the fixed points of some suitable primitive substitutions, which
means the generalized kolakoski sequences are uniformly recurrent.

    In this paper, our main goal is to study the corresponding
problems on $n$-letter alphabets having same remainder $r$ when
divided by $n$. By using distinct methods, we give all the fixpoints
of the operator $\Delta$ (Theorem~\ref{t0}) and establish that (1)
smooth infinite words are recurrent (Theorem~\ref{thm1}); (2) if the
remainder $r$ is 0  or $r>0$ and $n$ is an even number, the
generalized Kolakoski words are uniformly recurrent for $n$-letter
alphabets with the cyclic order (Theorem~\ref{thmm}); (3) if $r=0$,
the letters frequency is $1/n$ for two times differentiable
well-proportioned words (Theorem~\ref{t2}), which means that the
frequency of  the generalized Kolakoski sequences is $1/2$ for
2-letter even alphabets (Corollary~\ref{c1}); (4) the factor set of
three times differentiable infinite words is not closed under any
nonidentical permutation (Theorem \ref{t1}). Moreover we provide a
new proof of smooth infinite words being closed under reversal for
odd alphabets.

The paper is organized as follows. In Section \ref{s2}, we shall
first fix some notations and introduce some notions. Secondly in
Section \ref{s3}, we give all the fixpoints of the operator $\Delta$
over $\Sigma_n$. In Section \ref{s4}, we establish the frequency of
two times differentiable well-proportioned words for $n$-letter
alphabets having the remainder $r=0$. In Section \ref{s10}, we show
that three times differentiable infinite words  are not closed under
any nonidentical permutation for $n$-letter alphabets. In Section
\ref{s5}, we give some useful notations of the inverse $\Phi^{-1}$
of the bijection $\Phi:\mathcal{C}^{\omega}_n\rightarrow
\mathbf{\Sigma}_n^{\omega}$. In Section \ref{s6}, we establish some
lemmas which are used in the following discussions. In Section
\ref{s7}, we prove that smooth infinite words are reccurent for
$n$-letter alphabet with the cyclic order. In Section \ref{s8}, we
establish that if the remainder $r$ is 0, or $r>0$ and $n$ is an
even number, then the generalized Kolakoski sequence is the fixpoint
of some suitable primitive substitution for $n$-letter alphabets
having same remainder $r$ with the cyclic order, which means that
the generalized Kolakoski sequence is uniformly reccurent. In
Section \ref{s9}, we give a new proof of smooth infinite words being
closed under reversal for 2-letter odd alphabets. Finally we end
this paper with some concluding remarks in Section \ref{s11}.
\section{Definitions and notation\label{s2}}
Let $\Sigma_n=\{a_1,a_2,\cdots,a_n\}$, where $a_1,a_2,\cdots,a_n$
are positive integers with $a_1<a_2<\cdots<a_n$, $\Sigma^{*}_n$
denotes the free monoid over $\Sigma_n$. A \textit{finite word} over
$\Sigma_n$ is a member of $\Sigma_n^{*}$. If $w=w_{1}w_{2}\cdots
w_{k}$, $w_{i}\in \Sigma_n$ for $i=1, 2, \cdots, k$ then \emph{k} is
called the \textit{length} of the word \textit{w} and is denoted by
$|w|$. Sometimes we also use $w[i]$ to denote the $i$th letter $w_i$
of the word $w$, that is, $w=w[1]w[2]\cdots w[k]$, and $w[i\ldots
j]=w[i]w[i+1]\cdots w[j]$ for $1\leq i\leq j\leq k$. If $|w|$=0 then
\textit{w} is called the \textit{empty word} and is denoted by
$\varepsilon$. Let $|w|_\alpha$ be the number of $\alpha$ which
occurs in $w$, where $\alpha\in \Sigma_n$, then
$|w|=\Sigma_{i=1}^n|w|_{a_i}$. Moreover $\Sigma_n^{+}$ denotes
$\Sigma_n^{*}-\{\varepsilon\},\;\Sigma_n^k=\{w\in\Sigma_n^{*}:\;
|w|=k\}$ for $k=0,1,2,\cdots$.

  The set of all right infinite words is denoted by $\Sigma_n^{\omega}$, and
$\Sigma_n^{\infty}=\Sigma_n^{\omega}\cup \Sigma_n^{*}$. Given a word
$w\in\Sigma_n^{\infty}$, a \textit{factor} \emph{u} of \emph{w} is a
word $\emph{u}\in \Sigma_n^\infty$ such that \emph{w=xuy} for
$x,y\in \Sigma_n^{\infty}$, and $F(w)$ denotes the set of all
factors of $w$, $F_k(w)=F(w)\cap \Sigma_n^k$.
   If $x=\varepsilon$ (resp. $y=\varepsilon)$ then \emph{u} is
called \textit{prefix} (resp. \textit{suffix)}. A \textit{run} (or
\textit{block}) is a maximal factor of form $u=\alpha^{k},
\alpha\in\Sigma_n$. $\mathit{Pref(w)}$ denotes the set of all
prefixes of \emph{w}, $\mathit{Pref_n(w)}$ denotes the prefix of
length $n$ of $w$. Finally $N$ denotes the set of all natural
numbers, $\mathcal{N}^{*}\text{ and}\;\mathcal{N}^{\omega}$ denote
the free monoid and the set of all right infinite words over
$\mathcal{N}$ respectively, where $\mathcal{N}$ is the set of
positive integers.

   Let $u=u_{1}u_{2}\cdots u_k\in\Sigma_n^{*}$, where $u_{i}\in \Sigma_n,\; i=1,2,\cdots  u_k$.
The \textit{reversal} of $u$ is the word $\tilde{u}= u_k \cdots
u_{2}u_{1}$. A \textit{palindrome} is a word \emph{P} such that
$P=\tilde{P}$. Let $u=u_{1}u_{2}\cdots\in\Sigma_n^{\infty}$, where
$u_{i}\in \Sigma_n,\; i=1,2,\cdots$. The \textit{permutation}
$\sigma(u)$ of $u$ is the word
$\sigma(u)=\sigma(u_{1})\sigma(u_{2})\cdots $, where $\sigma$ is a
permutation of $\Sigma_n$. Especially, if $n=2$, then $\Sigma_2$ has
an unique nonidentical permutation  $\bar{ }$, which is determined
by $\bar{a}_1=a_2,\; \bar{a}_2=a_1$ and is also said to be the
\textit{complement}.

An infinite word $w$ is \textit{recurrent} if every factor has
infinitely many occurrences. And an infinite word $w$ is
\textit{uniformly recurrent} if every factor appears infinitely
often with bounded gap.

We see that every word $\emph{w}\in\Sigma_n^{\infty}$ can be
exclusively written as a product of factors as follows:
\begin{eqnarray}
w=\alpha_1^{i_{1}}\alpha_2^{i_{2}}\alpha_3^{i_{3}}\cdots,\,\text{where}\,\alpha_j\in \Sigma_n,\,\alpha_j\neq \alpha_{j+1}\, \text{and}\,i_{j}\in \mathcal{N}\cup \{\omega\} \text{ for } j\in \mathcal{N}.\label{eq2.1}
\end{eqnarray}
From  (\ref{eq2.1}) it is obvious that $w$ is uniquely determined by
the sequences $\prod_{j\geq 1} i_j$ and $\prod_{j\geq 1} \alpha_j$,
which are respectively said  to be the \textit{base sequence} and
\textit{power index sequence} of $w$, and be respectively denoted by
$\Delta(w)$ and $\mathcal{B}(w)$. That is
\begin{eqnarray}
  &\Delta:&\Sigma_n^{*}\rightarrow \mathcal{N}^{*}, \text{ defined
  by }\nonumber\\
 &\Delta(w)&=i_{1}i_{2}i_{3}\cdots=\prod_{k\geq 1}i_{k};\label{eq2.2}\\
  &\mathcal{B}:&\Sigma_n^{*}\rightarrow \Sigma_n^{*}, \text{ defined
  by }\nonumber\\
  &\mathcal{B}(w)&=\alpha_{1}\alpha_{2}\alpha_{3}\cdots=\prod_{k\geq
  1}\alpha_{k},\label{eq2.3}
\end{eqnarray}
which is easily extended to $\Sigma_n^\omega$ (also refers to
\cite{Br1}). From (\ref{eq2.1}-\ref{eq2.3}) it immediately follows
that
\begin{eqnarray}
w\in \Sigma_n^* \text{ is a palindrome}\Longleftrightarrow
\Delta(w)\text{ and }\mathcal{B}(w)\text{ are both
palindromes}.\label{eq2.4}
\end{eqnarray}
The function $(\Delta,\,\mathcal{B})$ gives the run-length coding
(\textit{RLE}) on $\Sigma_n^*$, which is a very simple form of data
compression in which runs of data are stored as a single data value
and count, rather than as the original run. Run-length encoding
performs lossless data compression and is well suited to
palette-based iconic images. For example, it is used in fax machines
and is relatively efficient because most faxed documents are mostly
white space, with occasional
interruptions of black.\\
{\bf Remark 1:} If $w\in \Sigma_n^{\omega}$ in  (\ref{eq2.1}) and
$|\mathcal{B}(w)|=k$, then $\alpha_k=\omega$. For example,
$w=2^34^23^44^\omega\in \{2,3,4\}^\omega$, then
$\Delta(w)=324\omega$ and $\mathcal{B}(w)=2434$.

  For any $\emph{w}\in \Sigma_n^{\infty}$, $\mathit{first(w)}$
denotes the first letter of the word \emph{w}. For each $\emph{w}
\in \Sigma_n^{*}$, $\mathit{last(w)}$ denotes the last letter of the
word \emph{w}. It is clear that the operator $\Delta$ satisfies the
property:
   $\Delta(uv)=\Delta(u)\Delta(v)$ if and only if  $\mathit{last(u)}\neq
   \mathit{first(v)}$.

Since $\Delta(w)$ is independent of the choice of the base sequence
$\mathcal{B}(w)$, the function $\Delta$ is not injective because
$\Delta(\sigma w)=\Delta(w)$ for every word \emph{w} and any
permutation $\sigma$ of $\Sigma_n$. For instance, if
$w=2^21^33^57^6(4^35^26^4)^{1000}$ then $\Delta(w)=2356(324)^{1000}$
and $\mathcal{B}(w)=2137(456)^{1000}$. In fact, over
$\Sigma_7=\{1,2,3,4,5,6,7\}$, there are $7\cdot 6^{1003}$ different
choices of the base sequence $\mathcal{B}(w)$ having the same power
exponent sequence $2356(324)^{1000}$ with the word $w$. In general,
for each word $w$ over $\Sigma_n\;(n\geq 2)$, if $|\Delta(w)|=k$,
then there are $n(n-1)^{k-1}$ different words over $\Sigma_n$ having
the same power index sequence $\Delta(w)$ with the word $w$.

Let $\mathfrak{B}_n=\{u\in \Sigma_n^\omega:\,u_i\neq u_{i+1}\text{
for } i\in \mathcal{N}\}$. If $n=2$ then
$\mathfrak{B}_2=\{(a_1a_2)^\omega,(a_2a_1)^\omega\}$; if $n>2$ then
$|\mathfrak{B}_n|=\infty$. Thus, given a sequence $B\in
\mathfrak{B}_n$, the \textit{pseudo-inverse function}

  $\;\;\;\Delta_{B}^{-1}:
\Sigma_n^{\infty}\rightarrow\Sigma_n^{\infty},\; u=u_{1}u_{2}u_{3}u_{4}\cdots$\\
can be defined by
\begin{eqnarray}\label{eq2.5}
\Delta_{B}^{-1}(u)=\left\{\begin{array}{ll}
b_1^{u_{1}}b_2^{u_{2}}b_3^{u_{3}}\cdots,& \;\textrm{ if }u\in \Sigma^\omega;\\
b_1^{u_{1}}b_2^{u_{2}}\cdots b_k^{u_k}, & \;\textrm{ if } |u|=k.
\end{array}\right.
\end{eqnarray}
Then the following property is immediate:

$\forall u\in \Sigma_n^{\infty}: \sigma
\Delta^{-1}_B(u)={\Delta^{-1}_{\sigma B}(u)}$, where $\sigma$ is a
permutation of  $\Sigma_n$; \vspace{.15cm}

$\forall u\in \Sigma_n^{*}:
\widetilde{\Delta^{-1}_{B}(\tilde{u})}=\Delta^{-1}_{\widetilde{B}}(u)$.

Now we first generalize the notion of finite differentiable word
over the alphabet $\{1,\,2\}$, which was given first by
Dekking~\cite{Dekking2}, to the one of infinite differentiable word
over any $n$-letter alphabet $\Sigma_n$.
\begin{defn}\label{def1} \rm
Let $w=\alpha_1^{t_{1}}\alpha_2^{t_{2}}\cdots \in \Sigma_n^{\omega}$
and $k$ be a fixed positive integer.

(1) If $\Delta^{i}(w)\in \Sigma_n^{\omega}$ for $i=1,2,\cdots,k$,
then we call $w$ a \textit{$k$-times differentiable infinite word}
over $\Sigma_n$;

(2) If $\Delta^{k}(w)\in \Sigma_n^{\omega}$ for any positive integer
$k$ then we call $w$ a \textit{$C_n^{\omega}$-word} (or a
\textit{smooth infinite word}) over $\Sigma_n$.
\end{defn}
For example, let
$w=(3^{3}1^{3}3^{3}1313^{3}1^{3}3^{3}13^{3}13^{3}1^{3}3^{3}1313^{3}1^{3}3^{3}13^{3}1^{3}3^{3}1)^{\omega}$
then $w$ is a 4-times differentiable word, but it is not a 5-times
differentiable word over $\{1,3\}$.

Secondly we extend the notion of differentiable words over the
alphabet $\{1,2\}$ to over arbitrary $n$-letters alphabet
$\Sigma_n$.
\begin{defn}\label{def2}\rm
Let $w\in \Sigma_n^{*}$ and
$w=\alpha_1^{t_{1}}\alpha_2^{t_{2}}\cdots \alpha_k^{t_{k}},
\textrm{where} \;\alpha_i\in \Sigma_n \text{ and } \alpha_i\neq
\alpha_{i+1}\, \textrm{for} \,1\leq i\leq k-1,$  and $1\leq
t_{i}\leq a_n$ for $i=1,2,\cdots,k$.

 If $t_{i}\in \Sigma_n$ for $i=2,3,\cdots,k-1$, then we call that
$w$ is \textit{differentiable}, and its \textit{derivative}, denoted
by $D(w)$, is the word whose \emph{j}th symbol equals the
 length of the \emph{j}th run of \emph{w}, discarding the first
 and/or the last run if its length is less than $a_n$.
\end{defn}
Let $w\in \Sigma_n^{*}$ and $k$ be a fixed positive integer,  if
$D^{k}(w)\in \Sigma_n^{*}$ then we call $w$ a \textit{$k$-times
differentiable word} over $\Sigma_n$, or a \textit{$C^{k}_n$-word}.
If a finite word $w$ is arbitrarily often differentiable, then we
call $w$ a \textit{$C^{\infty}_n$-word}, or a \textit{smooth word}.
Obviously, if \emph{w} is a smooth word and $|w|>0$, then
$|D(w)|<|w|$. Moreover, it is clear that $D$ is an operator from
$\Sigma_n^{*}$ to $\Sigma_n^{*}$ and
\begin{eqnarray}\label{eq2.6}
D(w)=\left\{\begin{array}{ll}
\varepsilon ,& \Delta(w)=yz, \;\textrm{where}\;y+z\geq 1,\;y,\,z<a_n\;\,  \text{or} \; w= \varepsilon\\
\Delta(w),& \Delta(w)=a_nxa_n  \; \text{or} \; \Delta(w)= a_n\\
xa_n, & \Delta(w)=yxa_n \;\textrm{and}\; 1\leq y<a_n\\
a_nx, & \Delta(w)=a_nxz \;\textrm{and}\;1\leq z<a_n\\
x, & \Delta(w)=yxz \;\textrm{and}\; 1\leq y,\,z<a_n\\
\end{array}\right.
\end{eqnarray}

 It is easy to see that finite factors of smooth infinite words are all $C^{\infty}$-words.
Thus finite smooth words~\cite{Br1}, which are finite factors of
smooth infinite words, are always $C^{\infty}$-words.

Let $w=\alpha^{b-1}\overbrace{\bar{\alpha}^{b}\alpha^{b}\cdots
\beta^{b}}^{b}\bar{\beta}^{b-1}\in C^{\infty}$, where if $2\mid b$
then $\beta=\alpha$, otherwise $\beta=\bar{\alpha}$, if $b-a>1$ then
$w$ can not occur in any smooth infinite word, so $C^{\infty}$-words
are not always finite smooth words.

   In what follows, we use
$C_n^{k}$ and $C_n^{k\omega}$ to stand for the sets of all $k$-times
differentiable finite words and all $k$-times differentiable
infinite words respectively.

It is easy to check that $\Delta$ and $D$ all commute with the
mirror image ($\tilde{\,\,}$) and are stable for every permutation
$\sigma$ over the alphabet $\Sigma_n$. Thus Proposition 4
in~\cite{Br4} still holds for arbitrary $n$-letter alphabets.
\begin{lem}\label{lem1} \rm Let $\sigma$ be a permutation of
$\Sigma_n$, then

(1) For all \emph{u} $\in\Sigma_n^{*},\;
D(\tilde{u})=\widetilde{D(u)},\; D(\sigma u)=D(u)$;

(2) For all \emph{u} $\in \Sigma_n^{\infty},\;
\Delta(\tilde{u})=\widetilde{\Delta(u)},\; \Delta(\sigma
u)=\Delta(u)$ (~\cite{Br1} Proposition 1 (a)-(b)). $\Box$
\end{lem}
\noindent These properties indicate that $C^{\infty}_n,\;
C^{\omega}_n,\;C^{k}_n\;\text{and}\;C^{k\omega}_n$ are all closed
under these operators:

  $w \in C^{\infty}_n\Longleftrightarrow\sigma w, \tilde{w}\in
C^{\infty}_n$;

  $w\in C^{\omega}_n\Longleftrightarrow\sigma w\in
C^{\omega}_n$;

  $w \in C^{k}_n\Longleftrightarrow\sigma w, \tilde{w}\in
C^{k}_n$;

  $w\in C^{k\omega}_n\Longleftrightarrow\sigma w\in C^{k\omega}_n$.
\section{Fixpoints of the operator $\Delta$\label{s3}}
The operator $\Delta$ over $\Sigma_2^\omega\;(a_1=a,\,a_2=b)$ has
exactly two fixpoints, that is $\Delta(K_{a,b})=K_{a,b},\;
\Delta(K_{b,a})=K_{b,a}$, where $K_{a,b}\;(\textrm{or}\,K_{b,a})$ is
an infinite sequence over the alphabet $\Sigma_2$ which starts with
$a\;(\textrm{or}\,b)$ and equals the sequence defined by its run
lengths: if $a>1$ then

$K_{a,b}=\underbrace{a^{a}b^{a}\cdots\alpha^{a}}_{a}
\underbrace{\bar{\alpha}^{b}\alpha^{b}\cdots b^{b}}_{a}\cdots$,
where $\alpha=b$ if $a$ is even, otherwise $\alpha=a,$

$K_{b,a}=\underbrace{b^{b}a^{b}\cdots\beta^{b}}_{b}
\underbrace{\bar{\beta}^{a}\alpha^{a}\cdots a^{a}}_{b}\cdots$,
where $\beta=a$ if $b$ is even, or else $\beta=b,$\\
which are called the generalized Kolakoski sequences.

If $n>2$, then the operator $\Delta$ over $\Sigma_n^{\omega}$ has
infinitely many fixpoints (see~\cite{Br1} Concluding remarks). In
fact, we can determine all the fixpoints of the operator $\Delta$.
To do so, let $\mathit{Fix(\Delta)}$ denote the set of all fixed
points of the operator $\Delta$ and $y=y_1^{t_1}
y^{t_2}_2y^{t_3}_3\cdots \in \mathit{Fix(\Delta)}$, where
$y_1y_2\cdots y_i\cdots\in \mathfrak{B}_n$. Then by $y=\Delta(y)$,
we have $t_1=t_2=\cdots=t_{y_1}=y_1$,
$t_{y_1+1}=t_{y_1+2}=\cdots=t_{y_1+y_2}=y_2\cdots$, $t_{(y_1-1)\cdot
y_1+1}=t_{(y_1-1)\cdot y_1+2}=\cdots=t_{y_1\cdot
y_1}=y_{y_1},\;\cdots$, it immediately follows that
\begin{thm}\label{t0}\rm
Let $\Sigma_n=\{a_1,a_2,\cdots,a_n\}$, where $a_1<a_2<\cdots<a_n\in
\mathcal{N}$. Then $\mathit{Fix(\Delta)}=\{K_u:\;u\in
\mathfrak{B}_n\}$ and
\begin{eqnarray}
K_u=u_1^{u_1}u_2^{u_1}\cdots
u_{u_1}^{u_1}u_{u_1+1}^{u_2}u_{u_1+2}^{u_2}\cdots
u_{u_1+u_1}^{u_2}\cdots u_{(u_1-1)\cdot u_1+1}^{u_{u_1}}\cdots
u_{u_1\cdot u_1}^{u_{u_1}}\cdots ,\label{eq1.1}
\end{eqnarray}
where $u=u_1u_2u_3\cdots$. $\Box$
\end{thm}
The sequences satisfying  (\ref{eq1.1}) are  said to be the
generalized Kolakoski sequences. Theorem~\ref{t0} suggests that if
$n>2$ then there are infinitely many generalized Kolokoski sequences
over the $n$-letters alphabet $\Sigma_n$. Moreover, it is obvious
that the generalized Kolakoski word $K_u$ over $\Sigma_n$ is a
smooth infinite word for every $u\in \mathfrak{B}_n$.
\section{Letter frequencies\label{s4}}
Keane~\cite{Keane} put forward whether the density of $1'$s in $K$
is 0.5. It is still a challenging problem. Actually, the best upper
density of $1'$s as well as the upper density of $2'$s in $K$ is
0.50084 and is due to Chv\'{a}tal~\cite{Chv}. Brlek, Jamet and
Paquin~\cite{Br3} showed that the frequency of letters in extremal
words is 1/2 for even alphabets, and for $a = 1$ with $b$ odd, the
frequency of $b$'s is $1/(\sqrt{2b-1}+ 1)$. Hereinafter we establish
the frequency of letters in 2-times differentiable well-proportioned
infinite words for arbitrary $n$-letters alphabets with each member
being a multiple of $n$.
\begin{lem}\label{llem2}\rm
Let $u=\alpha_1^{t_{1}}\alpha_2^{t_{2}}\cdots \alpha_k^{t_{k}}$,
where $\alpha_i,\;t_{i}\in \Sigma_n=\{a_1,a_2,\cdots,a_n\}$, $1\leq
i\leq k$, and $n\mid a_j$ for $j=1,2,\cdots,n$. And let
$m=|u|=\Sigma_{i=1}^k t_i$, then for any $v=v_1v_2\cdots v_m\in
\Sigma_n^*$ such that $\{v_{i\cdot n+1},v_{i\cdot
n+2},\cdots,v_{i\cdot n+n}\}=\{a_1,a_2,\cdots,a_n\}$, where
$i=0,1,\cdots, p$ and $m=n\cdot p$, one has
\begin{eqnarray}
|\Delta^{-m}_{v}(u)|_{a_i}
                             =\frac{|\Delta^{-m}_{v}(u)|}{n}
                             \text{ for } i=1,2,\cdots,n. \label{eq4.1}
\end{eqnarray}
\end{lem}
\prf{Since $t_{1},\,t_{2},\,\cdots, t_{k}\in \Sigma_n$ and $n\mid
a_i$, we see that $t_{i}=n\cdot s_i$ for $i=1,2,\cdots,k$. Thus
\begin{eqnarray}
\Delta^{-m}_{v}(u)&=&\overbrace{v_1^{\alpha_1}v_2^{\alpha_1}\cdots
v_n^{\alpha_1}v_{n+1}^{\alpha_1}v_{n+2}^{\alpha_1}\cdots
v_{n+n}^{\alpha_1}\cdots
v_{(s_1-1)n+1}^{\alpha_1}v_{(s_1-1)n+2}^{\alpha_1}\cdots
v_{(s_1-1)n+n}^{\alpha_1}}^{t_{1}}\nonumber\\
{}&&\overbrace{v_{t_1+1}^{\alpha_2}v_{t_1+2}^{\alpha_2}\cdots
v_{t_1+n}^{\alpha_{2}}v_{t_1+n+1}^{\alpha_2}v_{t_1+n+2}^{\alpha_2}\cdots
v_{t_1+2\cdot n}^{\alpha_2}\cdots v_{q-n+1}^{\alpha_2}
v_{q-n+2}^{\alpha_2}\cdots v_{q}^{\alpha_2}}^{t_2}\nonumber\\
{}&&\cdots\nonumber\\
{}&&\overbrace{v_{h+1}^{\alpha_k}v_{h+2}^{\alpha_k}\cdots
v_{h+n}^{\alpha_{k}}v_{h+n+1}^{\alpha_k}v_{h+n+2}^{\alpha_k}\cdots
v_{h+2\cdot n}^{\alpha_k}\cdots v_{m-n+1}^{\alpha_k}
v_{m-n+2}^{\alpha_k}\cdots
v_{m}^{\alpha_k}}^{t_{k}}.\nonumber\\\label{eq4.2}
\end{eqnarray}
where $q=t_1+t_2,\, h=t_1+t_2+\cdots+t_{k-1}$. Since $n\mid t_j$ for
$j=1,2,\cdots,n$ and  $\{v_{i\cdot n+1},v_{i\cdot
n+2},\cdots,v_{i\cdot n+n}\}=\{a_1,a_2,\cdots,a_n\}$ for
$i=0,1,\cdots, p$, we easily see that there are equal numbers of
$a_1^{\alpha_i},a_2^{\alpha_i},\cdots,$ and $a_n^{\alpha_i}$ in the
subsequence
\\$$\overbrace{v_{b_i+1}^{\alpha_i}v_{b_i+2}^{\alpha_i}\cdots
v_{b_i+n}^{\alpha_{i}}v_{b_i+n+1}^{\alpha_i}v_{b_i+n+2}^{\alpha_i}\cdots
v_{b_i+2\cdot n}^{\alpha_i}\cdots v_{b_i+t_k-n+1}^{\alpha_i}
v_{b_i+t_k-n+2}^{\alpha_i}\cdots v_{b_i+t_k}^{\alpha_i}}^{t_{i}}$$
\\for $i=1,2,\cdots,k$, where $\alpha_i\in \Sigma_n$, $b_1=0$ and
$b_i=t_1+t_2+\cdots+t_{i-1}$ for $i=2,3,\cdots,k$. Therefore there are equal numbers of
$a_1^{\alpha_i},a_2^{\alpha_i},\cdots,$ and $a_n^{\alpha_i}$ in the
right side of  (\ref{eq4.2}), which suggest that  (\ref{eq4.1})
holds.}  $\Box$\\
{\bf Remark 2:} From the proof of Lemma~\ref{llem2}, we easily see that
Lemma~\ref{llem2} also holds if the condition $t_i\in \Sigma_n$ is
substituted by $t_i$ being divided by $n$.
\begin{defn}\label{def3}\rm
Let $w\in \Sigma_n^\omega$. If the base sequence
$\mathcal{B}(w)=\alpha_1\alpha_2\cdots\alpha_k\cdots$\; satisfies
\begin{eqnarray}
\{\alpha_{i\cdot n+1},\alpha_{i\cdot n+2},\cdots,\alpha_{i\cdot
n+n}\}=\{a_1,a_2,\cdots,a_n\}\text{ for }\forall\,i\in
\mathcal{N},\label{eq4.3}
\end{eqnarray}
then $w$ is said to be \textit{well-proportioned}.
\end{defn}

Now we are in a position to prove our second main result.
\begin{thm}\label{t2}\rm
Let $w$ be a two times differentiable well-proportioned infinite
word over the $n$-letters alphabet
$\Sigma_n=\{a_1,a_2,\cdots,a_n\}$. If $n\mid a_j$ for
$j=1,2,\cdots,n$, then
\begin{eqnarray}
\lim_{k\rightarrow \infty} \frac{|W_k|_{a_i}}{|W_k|}
                             =\frac{1}{n} \text{ for } i=1,2,\cdots,n,\label{eq4.4}
\end{eqnarray}
\end{thm}
\noindent where $\mathit{W_k=w[1\cdots k]=Pref_k(w)}$.

\prf{Since $\Delta^2(w)\in \Sigma_n^\omega$ we have
\begin{eqnarray}
\Delta(w)=\gamma_{1}^{t_1}\gamma_{2}^{t_2}\cdots
\gamma_{r}^{t_r}\cdots, \text{ where } \gamma_i,\;t_i\in \Sigma_n
\text{ for } i=1,2,\cdots.\label{eq4.5}
\end{eqnarray}
Thus  (\ref{eq4.5}) suggests
\begin{eqnarray}
w=\lim_{i\rightarrow
\infty}\Delta_{\mathcal{B}(w)_i}^{-1}(\Delta(w)_i), \label{eq4.6}
\end{eqnarray}
where
$\mathcal{B}(w)_i=\mathit{Pref_i(}\mathcal{B}(w)),\;\Delta(w)_i=\mathit{Pref_i(}\Delta(w))$.
Therefore from (\ref{eq4.6}) we see that for each $k\in
\mathcal{N}$, there is a positive integer $i$ such that
\begin{eqnarray}
W_k=\Delta_{\mathcal{B}(w)_i}^{-1}(\Delta(w)_i)v, \text{ where }v\in
\Sigma_n^* \text{ and } |v|\leq a_{n}^{a_n}.\label{eq4.7}
\end{eqnarray}
Now  (\ref{eq4.3}), (\ref{eq4.7}) and (\ref{eq4.1}) imply that
 (\ref{eq4.4}) holds.}   $\Box$

Since infinite words over 2-letter alphabets are always
well-proportioned, from Theorem \ref{t2} we can deduce
\begin{cor}\rm\label{c1}
The frequency of letters in 2-times differentiable infinite words is
1/2 for 2-letter even alphabets.  $\Box$
\end{cor}
Furthermore from Theorem \ref{t2} it immediately follows that
\begin{cor}\label{c2}\rm
If $n\mid a_i$ for $i=1,2,\cdots,n$, then the frequency of letters
in well-proportioned smooth infinite words is $1/n$ for the alphabet
$\Sigma_n$.  $\Box$
\end{cor}
\section{Permutation\label{s10}}
Kimberling~\cite{Kimberling} first raised whether or not the
complement of a finite factor of Kolakoski sequence $K$ is still a
factor of $K$ and whether or not every finite factor of $K$ occurs
infinitely often (recurrence). Dekking~\cite{Dekking2} showed that
the closure of $F(K)$ by complementation would imply the recurrence
property. These conjectures were stated for every smooth infinite
word over $\{1, 2\}$ in~\cite{Br2}. Brlek etc.~\cite{Br4} proved
that the existence of arbitrarily long palindromes in infinite
smooth words on $\{1, 2\}$ would imply the recurrence property.
Recently Brlek etc.~\cite{Br3} showed that the closure of $F(w)$
under reversal holds for odd alphabets only (\cite{Br3} Proposition
15) and that the set of factors of extremal smooth words is not
closed under reversal and under complementation (\cite{Br3}
Proposition 26) for even alphabets. Corollary 9 (iii) in~\cite{Br3}
implies that the set of factors of extremal smooth words on
$\{1,3\}$ is not closed under complementation. Next we establish the
corresponding result for $n$-letter alphabets having same remainder
when divided by $n$. And the proof seems to be slightly more
essential and straightforward.
\begin{thm}\label{t1}\rm
Let $\Sigma_n=\{a_1,a_2,\cdots,a_n\}$, where $a_1<a_2<\cdots<a_n\in
\mathcal{N}$ and  $a_i\equiv r\,(\text{mod }n)$ for
$i=1,2,\cdots,n$, where $r=0,1,\cdots,n-1$. Suppose that $w\in
C^{3\omega}_n$ and $\mathcal{B}(w)=(b_1b_2\cdots
b_n)^\omega,\;\mathcal{B}(\Delta(w))=(c_1c_2\cdots
c_n)^\omega,\;\mathcal{B}(\Delta^2(w))=(e_1e_2\cdots e_n)^\omega$,
where $b_1b_2\cdots b_n,\,c_1c_2\cdots c_n\text{ and }e_1e_2\cdots
e_n$ are arrangements of $a_1,a_2,\cdots,a_n$. Then $F(w)$ is not
closed under any nonidentical permutation.
\end{thm}
\prf{Since $w\in C^{3\omega}_n$ and
$\mathcal{B}(\Delta^2(w))=(e_1e_2\cdots e_n)^\omega$,  we have
\begin{equation}\label{eq5.1}
\Delta^2(w)=e^{t_1}_1e^{t_2}_2\cdots
e^{t_n}_ne^{t_{n+1}}_1e^{t_{n+2}}_2\cdots e^{t_{2n}}_n\cdots
e^{t_{k\cdot n+1}}_1e^{t_{k\cdot n+2}}_2\cdots e^{t_{(k+1)\cdot
n}}_n\cdots
\end{equation}
where $t_i\in \Sigma_n$ for $i\in \mathcal{N}$.

From $e_i,\,t_i\in \Sigma_n$, we get $e_i=n\cdot q_i +r,\,q_i\in
N\text{ for }i=1,2,\cdots,n;\;t_i=n\cdot h_i+ r,h_i\in N\text{ for
}i=1,2,\cdots$, where $r=0,1,\cdots,n-1$. Now let $c_{n\cdot
k+i}=c_i$ for $i=1,2,\cdots,n$ and $k\in \mathcal{N}$, then note
that $c_{n\cdot k}=c_n$, by $\mathcal{B}(\Delta(w))=(c_1c_2\cdots
c_n)^\omega$ and (\ref{eq5.1}), we can obtain
\begin{eqnarray}
 \Delta(w)&=&(c^{e_1}_1c^{e_1}_2\cdots c^{e_1}_n)^{h_1}c^{e_1}_1c^{e_1}_2\cdots
c^{e_1}_r(c^{e_2}_{r+1}c^{e_2}_{r+2}\cdots
c^{e_2}_r)^{h_2}c^{e_2}_{r+1}c^{e_2}_{r+2}\cdots c^{e_2}_{2r}\cdots\nonumber\\
{}&&(c^{e_n}_{(n-1)r+1}c^{e_n}_{(n-1)r+2}\cdots c^{e_n}_{(n-1)
r})^{h_n}c^{e_n}_{(n-1)r+1}c^{e_n}_{(n-1)r+2}\cdots c^{e_n}_{n\cdot
r}\nonumber
\end{eqnarray}
\begin{eqnarray}
{}&&\hspace{1.2cm}(c^{e_1}_1c^{e_1}_2\cdots
c^{e_1}_n)^{h_{n+1}}c^{e_1}_1c^{e_1}_2\cdots
c^{e_1}_r(c^{e_2}_{r+1}c^{e_2}_{r+2}\cdots
c^{e_2}_r)^{h_{n+2}}c^{e_2}_{r+1}c^{e_2}_{r+2}\cdots c^{e_2}_{2r}\cdots\nonumber\\
{}&&\hspace{1.2cm}(c^{e_n}_{(n-1)r+1}c^{e_n}_{(n-1)r+2}\cdots c^{e_n}_{(n-1)
r})^{h_{2n}}c^{e_n}_{(n-1)r+1}c^{e_n}_{(n-1)r+2}\cdots
c^{e_n}_{n\cdot r}\cdots\label{eq5.2}
\end{eqnarray}
Let $b_{n\cdot i+j}=b_j$ for $j=1,2,\cdots,n$ and $i\in
\mathcal{N}$, then note that $b_{n\cdot i}=b_n$, from
$\mathcal{B}(w)=(b_1b_2\cdots b_n)^\omega$ and (\ref{eq5.2}) it
immediately follows that
\begin{eqnarray}\label{eq5.3}
w&=&[(b^{c_1}_1b^{c_1}_2\cdots
b^{c_1}_n)^{q_1}b^{c_1}_1b^{c_1}_2\cdots b^{c_1}_r
(b^{c_2}_{r+1}b^{c_2}_{r+2}\cdots
b^{c_2}_{r})^{q_1}b^{c_2}_{r+1}b^{c_2}_{r+2}\cdots b^{c_2}_{2\cdot
r}\cdots\nonumber\\
{}&&(b^{c_n}_{(n-1)r+1}b^{c_n}_{(n-1)r+2}\cdots
b^{c_n}_{(n-1)r})^{q_1}b^{c_n}_{(n-1)r+1}b^{c_n}_{(n-1)r+2}\cdots
b^{c_n}_{n\cdot r}]^{h_1}\nonumber\\
{}&&(b^{c_1}_1b^{c_1}_2\cdots
b^{c_1}_n)^{q_1}b^{c_1}_1b^{c_1}_2\cdots b^{c_1}_r
(b^{c_2}_{r+1}b^{c_2}_{r+2}\cdots
b^{c_2}_{r})^{q_1}b^{c_2}_{r+1}b^{c_2}_{r+2}\cdots
b^{c_2}_{2\cdot r}\cdots \nonumber\\
{}&&(b^{c_r}_{(r-1)r+1}b^{c_r}_{(r-1)r+2}\cdots
b^{c_r}_{(r-1)r})^{q_1}b^{c_r}_{(r-1)r+1}b^{c_r}_{(r-1)r+2}\cdots
b^{c_r}_{r^2}
\nonumber\\
{}&& [(b^{c_{r+1}}_{r^2+1}\cdots
b^{c_{r+1}}_{r^2})^{q_2}b^{c_{r+1}}_{r^2+1}\cdots
b^{c_{r+1}}_{r^2+r} (b^{c_{r+2}}_{r^2+r+1}\cdots
b^{c_{r+2}}_{r^2+r})^{q_2}b^{c_{r+2}}_{r^2+r+1}\cdots
b^{c_{r+2}}_{r^2+2\cdot
r}\cdots\nonumber\\
{}&&(b^{c_r}_{r^2+(n-1)r+1}\cdots
b^{c_r}_{r^2+(n-1)r})^{q_2}b^{c_r}_{r^2+(n-1)r+1}\cdots
b^{c_r}_{r^2}]^{h_2}\nonumber\\
{}&&(b^{c_{r+1}}_{r^2+1}\cdots
b^{c_{r+1}}_{r^2})^{q_2}b^{c_{r+1}}_{r^2+1}\cdots
b^{c_{r+1}}_{r^2+r} (b^{c_{r+2}}_{r^2+r+1}\cdots
b^{c_{r+2}}_{r^2+r})^{q_2}b^{c_{r+2}}_{r^2+r+1}\cdots
b^{c_{r+2}}_{r^2+2\cdot
r}\cdots\nonumber\\
{}&&(b^{c_{2r}}_{r^2+(r-1)r+1}\cdots
b^{c_{2r}}_{r^2+(r-1)r})^{q_2}b^{c_{2r}}_{r^2+(r-1)r+1}\cdots
b^{c_{2r}}_{2r^2}\cdots
\nonumber\\
{}&&[(b^{c_{(n-1)r+1}}_{(n-1)r^2+1}\cdots
b^{c_{(n-1)r+1}}_{(n-1)r^2})^{q_n}b^{c_{(n-1)r+1}}_{(n-1)r^2+1}\cdots
b^{c_{(n-1)r+1}}_{(n-1)r^2+r} \nonumber\\
{}&&(b^{c_{(n-1)r+2}}_{(n-1)r^2+r+1}\cdots
b^{c_{(n-1)r+2}}_{(n-1)r^2+r})^{q_n}b^{c_{(n-1)r+2}}_{(n-1)r^2+r+1}\cdots
b^{c_{(n-1)r+2}}_{(n-1)r^2+2\cdot r}\cdots \nonumber\\
{}&&(b^{c_{(n-1)r}}_{(n-1)r^2+(n-1)r+1}\cdots
b^{c_{(n-1)r}}_{(n-1)r^2+(n-1)r})^{q_n}b^{c_{(n-1)r}}_{(n-1)r^2+(n-1)r+1}\cdots
b^{c_{(n-1)r}}_{(n-1)r^2}]^{h_n}\nonumber\\
{}&&(b^{c_{(n-1)r+1}}_{(n-1)r^2+1}\cdots
b^{c_{(n-1)r+1}}_{(n-1)r^2})^{q_n}b^{c_{(n-1)r+1}}_{(n-1)r^2+1}\cdots
b^{c_{(n-1)r+1}}_{(n-1)r^2+r}\nonumber\\
{}&& (b^{c_{(n-1)r+2}}_{(n-1)r^2+r+1}\cdots
b^{c_{(n-1)r+2}}_{(n-1)r^2+r})^{q_n}b^{c_{(n-1)r+2}}_{(n-1)r^2+r+1}\cdots
b^{c_{(n-1)r+2}}_{(n-1)r^2+2\cdot
r}\cdots\nonumber\\
{}&&(b^{c_{nr}}_{(n-1)r^2+(r-1)r+1}\cdots
b^{c_{nr}}_{(n-1)r^2+(r-1)r})^{q_n}b^{c_{nr}}_{(n-1)r^2+(r-1)r+1}\cdots
b^{c_{nr}}_{nr^2}\cdots
\end{eqnarray}

From (\ref{eq5.2}, \ref{eq5.3}) we see that the maximal factors of
$w$, which are composed of consecutive runs of the same length, must
be one of the following forms.
\begin{eqnarray}
{}&&\overbrace{(b^{c_{(j-1)r+1}}_{(j-1)r^2+1}\cdots
b^{c_{(j-1)r+1}}_{(j-1)r^2+n})^{q_j}b^{c_{(j-1)r+1}}_{(j-1)r^2+1}\cdots
b^{c_{(j-1)r+1}}_{(j-1)r^2+r}}^{e_j},\nonumber\\{}&&
\overbrace{(b^{c_{(j-1)r+2}}_{(j-1)r^2+r+1}\cdots
b^{c_{(j-1)r+2}}_{(j-1)r^2+r})^{q_j}b^{c_{(j-1)r+2}}_{(j-1)r^2+r+1}\cdots
b^{c_{(j-1)r+2}}_{(j-1)r^2+2\cdot r}}^{e_j}, \nonumber\\
{}&& \hspace{4.5cm}
\vdots \nonumber\\
{}&& \overbrace{(b^{c_{(j-1)r+n}}_{(j-1)r^2+(n-1)r+1}\cdots
b^{c_{(j-1)r+n}}_{(j-1)r^2+(n-1)r})^{q_j}b^{c_{(j-1)r+n}}_{(j-1)r^2+(n-1)r+1}\cdots
b^{c_{(j-1)r+n}}_{(j-1)r^2+n\cdot r}}^{e_j},\label{eq5.4}
\end{eqnarray}
where $j=1,2,\cdots,n$.

Note that by (\ref{eq5.4}), the length of all possible maximal
factors of $w$, which is composed of consecutive runs of the same
length, is respectively equal to $e_j$ for $j=1,2,\cdots,n$. Since
$e_1,e_2,\cdots,e_n$ is an arrangement of $a_1,a_2,\cdots,a_n$,
there is some $k\in \mathcal{N}$ such that $e_k=a_n$. Thus on the
one hand, since $e_k=a_n$, taking $j=k$ in (\ref{eq5.4}), we get

\begin{eqnarray}
{}&&\overbrace{(b^{c_{(k-1)r+1}}_{(k-1)r^2+1}\cdots
b^{c_{(k-1)r+1}}_{(k-1)r^2+n})^{q_k}b^{c_{(k-1)r+1}}_{(k-1)r^2+1}\cdots
b^{c_{(k-1)r+1}}_{(k-1)r^2+r}}^{a_n},\nonumber\\{}&&
\overbrace{(b^{c_{(k-1)r+2}}_{(k-1)r^2+r+1}\cdots
b^{c_{(k-1)r+2}}_{(k-1)r^2+r})^{q_k}b^{c_{(k-1)r+2}}_{(k-1)r^2+r+1}\cdots
b^{c_{(k-1)r+2}}_{(k-1)r^2+2\cdot r}}^{a_n}, \nonumber\\
{}&& \hspace{4.5cm}
\vdots \nonumber\\
{}&& \overbrace{(b^{c_{(k-1)r+n}}_{(k-1)r^2+(n-1)r+1}\cdots
b^{c_{(k-1)r+n}}_{(k-1)r^2+(n-1)r})^{q_k}b^{c_{(k-1)r+n}}_{(k-1)r^2+(n-1)r+1}\cdots
b^{c_{(k-1)r+n}}_{(k-1)r^2+n\cdot r}}^{a_n}.\label{ee1}
\end{eqnarray}

On the other hand, since $a_n>a_j$ for $j<n$, from (\ref{eq5.4}) we
see that the maximal factors of the word $w$, which are composed of
consecutive runs of the same length and are of the greatest length,
have all occurred in (\ref{ee1}). Therefore, from
$c_{(k-1)r+1},c_{(k-1)r+2},\cdots,c_{(k-1)r+n}$ being an arrangement
of $a_1,a_2,\cdots,a_n$,  it follows that for any nonidentical
permutation $\sigma$ of $\Sigma_n$, the image of each member of
(\ref{ee1}) under the permutation $\sigma$ cannot occur in
(\ref{ee1}),  which means that $w$ is not closed under any
nonidentical permutation. $\Box$}

From the proof of Theorem \ref{t1}, we easily see that two times
differentiable infinite words are not closed under complementation
for 2-letter alphabets having same parity.
\section{The bijection $\Phi$ from $\mathcal{C}^{\omega}_n$ to $\Sigma_n^{\omega}$\label{s5}}
From now on, similar to~\cite{Br1} concluding remarks, we need to
assume a cyclic order $\mathbf{b_1b_2\cdots b_n}$ of the alphabet
$\Sigma_n$ such that the base sequence of every word is compatible
with this order, where $b_1,b_2,\cdots, b_n$ is a fixed arrangement
of the elements $a_1,a_2,\cdots,a_n$ of the alphabet $\Sigma_n$. And
suppose that $b_{t\cdot n+i}=b_{i}$ for any $t\in N$ and
$i=1,2,\cdots,n$.  Note that if $n=2$, then $\mathbf{a_1a_2}$ is the
unique cyclic order of $\Sigma_2$. In fact, there are exactly
$(n-1)!$ different cyclic orders of $\Sigma_n$. Let
\begin{eqnarray}
\mathcal{C}^{k}_n&=&\{w\in C^{k}_n: \mathcal{B}(w)\text{ is
compatible
with } \mathbf{b_1b_2\cdots b_n}\};\\
\mathcal{C}_n^{\infty}&=&\{w\in C^{\infty}_n: \mathcal{B}(w)\text{
is compatible with } \mathbf{b_1b_2\cdots b_n}\};\\
\mathcal{C}^{\omega}_n&=&\{w\in C^{\omega}_n: \mathcal{B}(w)\text{
is compatible with } \mathbf{b_1b_2\cdots b_n}\}.
\end{eqnarray}
{\bf Remark 3:} Note that
$\mathcal{C}^{k}_2=C^{k}_2,\;\mathcal{C}_2^{\infty}=C^{\infty}_2,\;\mathcal{C}^{\omega}_2=C^{\omega}_2$.

In what follows, the pseudo-inverse function
$\Delta_{\alpha}^{-1}\;(\alpha\in \Sigma_n)$ is compatible with the
cyclic order $\mathbf{b_1b_2\cdots b_n}$ of the alphabet $\Sigma_n$.
Let $p=\alpha_{1}\alpha_{2}\cdots \alpha_{k}$, where $\alpha_1,\alpha_2,\cdots,\alpha_k\\\in
\Sigma_n$, we use $\Delta_{p}^{-k}(u)$ to
denote $\Delta_{\alpha_{1}}^{-1}\Delta_{\alpha_{2}}^{-1}\cdots
\Delta_{\alpha_{k}}^{-1}(u)$. Clearly, if $p=p_{1}p_{2}$ then
$\Delta_{p}^{-|p|}(u)=\Delta_{p_{1}}^{-|p_{1}|}\Delta_{p_{2}}^{-|p_{2}|}(u)$.\\
For example, given a cyclic order $\mathbf{243}$ of the alphabet
$\{2,3,4\}$, then
\begin{eqnarray}
\Delta^{-1}_{2}(24)&=&2^24^4;\nonumber\\
\Delta^{-3}_{232}(24)&=&2^34^33^22^24^43^42^44^43^32^34^33^32^24^23^22^24^43^42^44^4\nonumber\\
                       &=&\Delta^{-1}_{2}\Delta^{-2}_{32}(24)\nonumber\\
                       &=&\Delta^{-2}_{23}\Delta^{-1}_{2}(24).\nonumber
\end{eqnarray}
In order to study infinitely often differentiable infinite words of
any period over $\{1,2\}$, Dekking~\cite{Dekking3} established a
bijection from $C^{\omega}_2$ to $\Sigma_2^{\omega}$. Similarly, we
can define a bijection from $\mathcal{C}^{\omega}_n$ to
$\Sigma_n^{\omega}$ as below:

\quad$\Phi:\mathcal{C}^{\omega}_n\longrightarrow\;\Sigma_n^{\omega}$\\
which is constructed by setting
\begin{eqnarray}
\Phi(w)[j+1]=\Delta^{j}(w)[1],\;\text{for}\; j\geq 0,\label{e3.1}
\end{eqnarray}
and its inverse is defined as follows:
\begin{eqnarray}
\Phi^{-1}(u)=\lim_{k\rightarrow \infty}\Delta^{-(k-1)}_{u[1\cdots
k-1]}(u[k]).\label{e3.2}
\end{eqnarray}
In the sequel, for any $u\in \Sigma^{\infty}$ and $1\leq i\leq |u|$,
$U_{i}$ denotes $u[1\cdots i]$ and
\begin{eqnarray}
\Phi^{-1}(u[1\cdots k])=\Delta^{-(k-1)}_{u[1\cdots
k-1]}(u[k]),\label{e3.3}
\end{eqnarray}
which determines a function from $\Sigma_n^+$ to $\Sigma_n^+$. Then
by (\ref{e3.2}-\ref{e3.3}), we get
\begin{eqnarray}
\Phi^{-1}(u)&=&\lim_{k\rightarrow\infty}\Phi^{-1}(u[1\cdots k])\label{e3.4}\\
            &=&\lim_{k\rightarrow\infty}\Phi^{-1}(U_{k}).\label{e3.5}
\end{eqnarray}
Obviously, $K_{b,a}=\Phi^{-1}(b^{\omega})$ and
$K_{a,b}=\Phi^{-1}(a^{\omega})$.
\section{Some Lemmas\label{s6}}
The following simple results are important in the sequel. First of
all, from the definition of the operator $\Delta_{\alpha}^{-1}
\;(\alpha\in \Sigma_n)$ it immediately follows that
\begin{lem}\label{lem2}\rm
Let $u\in \Sigma_n^{+}$, $v\in \Sigma_n^{+}\,(\Sigma_n^{\omega})$
and $\alpha\in \Sigma_n$.

(1) If $|u|=q\cdot n+k$, where $q\in \mathcal{N}$ and $k=0,1,\cdots,
n-1$, then

$\Delta_{\alpha}^{-1}(uv)=\Delta_{\alpha}^{-1}(u)\Delta_{b_j}^{-1}(v)$,
where $\alpha=b_i$ and $b_j=b_{i+k}$;

(2) If $n\mid |u|$, then

$\Delta_{\alpha}^{-1}(uv)=\Delta_{\alpha}^{-1}(u)\Delta_{\alpha}^{-1}(v)$.
$\Box$
\end{lem}
The following Lemma \ref{lem3} (1) is a generalization of Brlek et
al.~\cite{Br3} Lemma 21.
\begin{lem}\label{lem3}\rm
Let $w\in \Sigma_n^{+}$, $\alpha\in \Sigma_n$.

(1) If $a_1,a_2,\cdots,\text{ and }a_n$ when divided by $n$ have the
same remainder and $|w|$ is a multiple of $n$, then the length of
$\Delta_{\alpha}^{-1}(w)$ is also a multiple of $n$;

(2) If $n=2$, $a_1,\;a_2$ are odd integers and $w$ has odd length,
then $\Delta_{\alpha}^{-1}(w)$ also has odd length;

(3) If $n=2$, $a_1,\;a_2$ are odd integers and $w$ is a palindrome
of odd length, then $\Delta_{\alpha}^{-1}(w)$  is also a palindrome
of odd length.
\end{lem}
\prf{(1) Let $w=\alpha_{1}\alpha_{2}\cdots \alpha_{n\cdot k}$, where
$k\in \mathcal{N}$ and $\alpha_i\in \Sigma_n$ for $i\in
\mathcal{N}$, then
$\Delta_{\alpha}^{-1}(w)=\alpha^{\alpha_{1}}\bar{\alpha}^{\alpha_{2}}\cdots
\bar{\alpha}^{\alpha_{n\cdot k}}$. Thus since all
$\alpha_{1},\,\alpha_{2}\,\cdots,\,\text{ and }a_{n\cdot k}$ when
divided by $n$ have the same remainder, we have

$|\Delta_{\alpha}^{-1}(w)|=\sum\limits_{i=1}^{n\cdot k} \alpha_{i}=
\sum\limits_{i=1}^{n\cdot k} (n\cdot q_{i}+r)=n\cdot(
\sum\limits_{i=1}^{n\cdot k} q_{i}+k\cdot r)$,\\ where
$\alpha_{i}=n\cdot q_{i}+r,\;0\leq r \leq n-1$, which suggests that
the length of $\Delta_{\alpha}^{-1}(w)$ is a multiple of  $n$;

(2) Let $w=\alpha_{1}\alpha_{2}\cdots \alpha_{2k+1}$, where
$\alpha_{i}\in \Sigma_2$, then since
$\Delta_{\alpha}^{-1}(w)=\alpha^{\alpha_{1}}\bar{\alpha}^{\alpha_{2}}\cdots
\alpha^{\alpha_{2k+1}}$ and all $\alpha_{1},\,\alpha_{2},\,\cdots\,
\alpha_{2k+1}$ are odd integers, we see that
$|\Delta_{\alpha}^{-1}(w)|=\sum\limits^{2k+1}_{i=1} \alpha_{i}$ is
also an odd integer;

(3) Note that if $w$ has odd length, then it is clear that
$\Delta_{\alpha}^{-1}(\tilde{w})=\widetilde{\Delta_{\alpha}^{-1}(w)}$.
Thus if $w$ is a palindrome of odd length, then
$\Delta_{\alpha}^{-1}(w)$  is also a palindrome of odd length.}
$\Box$

From Lemmas \ref{lem2}-\ref{lem3}, it easily follows that
\begin{lem}\label{lem4}\rm
Let $w=uv$, $p=\alpha_{1}\alpha_{2}\cdots \alpha_{k}$, where $u\in
\Sigma_n^{+},\,v\in \Sigma_n^{+}\,(\Sigma^{\omega}),\;\alpha_{i}\in
\Sigma_n,\, i=1,2,\cdots k$.

(1) If $a_1,a_2,\cdots,\text{ and }a_n$ when divided by $n$ have the
same remainder and $|u|$ is a multiple of $n$, then

\quad$\Delta_{p}^{-k}(uv)=\Delta_{p}^{-k}(u)\Delta_{p}^{-k}(v)$;

(2) If $n=2,\;a_1,\;a_2$ are odd integers and $u$ has odd length,
then

\quad$\Delta_{p}^{-k}(uv)=\Delta_{p}^{-k}(u)\Delta_{\bar{p}}^{-k}(v)$.
$\Box$
\end{lem}
Furthermore from Lemma \ref{lem4} immediately follows:
\begin{lem}\label{lem5}\rm
Let $w=v_{1}v_{2}\cdots v_{n},\,p=\alpha_{1}\alpha_{2}\cdots
\alpha_{k}$, where $v_{j}\in \Sigma_n^{+},\,v_{n}\in
\Sigma_n^{+}\,\;(\Sigma_n^{\omega}),\,\alpha_{i}\in \Sigma_n,\,1\leq
j\leq n-1,\, 1\leq i\leq k$.

(1) If $a_1,a_2,\cdots\text{ and }a_n$ when divided by $n$ have the
same remainder and $|v_i|$ is a multiple of $n$ for $i=1,2,\ldots,
n-1$, then
\begin{equation}
\Delta_{p}^{-k}(w)=\Delta_{p}^{-k}(v_{1})\Delta_{p}^{-k}(v_{2})\cdots
\Delta_{p}^{-k}(v_{n});\label{e4.1}
\end{equation}

(2) If $n=2$ and $a_1,\;a_2$ are odd integers and all $v_{j}\;
(1\leq j\leq n-1)$ have odd length, then
\begin{equation}
\Delta_{p}^{-k}(w)=\Delta_{p}^{-k}(v_{1})\Delta_{\bar{p}}^{-k}(v_{2})\cdots
\Delta_{q}^{-k}(v_{n}),\label{e4.2}
\end{equation}
where if $2\mid n$ then $q=\bar{p}$, or else $q=p$.  $\Box$
\end{lem}

The function $\Phi^{-1}$ defined by (\ref{e3.3}) is of the following
important property, which guarantees that the limit of (\ref{e3.2})
exists.
\begin{lem}\label{lem6}\rm
If $u\in \mathit{Pref}(v),\;|u|=m$ and $|v|=n$, then
\begin{equation}
\Phi^{-1}(u)\in \mathit{Pref}(\Phi^{-1}(v))\label{e4.3}.
\end{equation}
\end{lem}
\prf{Since $u\in \mathit{Pref}(v)$, we have $v[1\cdots m]=u[1\cdots
m]$. Thus by  (\ref{e3.3}), we obtain
\begin{eqnarray}
\Phi^{-1}(v)&=&\Delta_{v[1\cdots n-1]}^{-(n-1)}(v[n])\nonumber\\
&=&\Delta_{v[1\cdots m-1]}^{-(m-1)}
\Delta_{v[m]}^{-1}\Delta_{v[m+1]}^{-1}\cdots\Delta_{v[n-1]}^{-1}(v[n])\nonumber\\
&=&\Delta_{u[1\cdots m-1]}^{-(m-1)}
(\Delta_{u[m]}^{-1}(\Delta_{v[m+1]}^{-1}\cdots\Delta_{v[n-1]}^{-1}(v[n])))\nonumber\\
&=&\Delta_{u[1\cdots m-1]}^{-(m-1)}(u[m]x)\nonumber\\
&=&\Delta_{u[1\cdots m-1]}^{-(m-1)}(u[m])y\nonumber\\
&=&\Phi^{-1}(u)y,\nonumber
\end{eqnarray}
which suggests that  (\ref{e4.3})  holds.}  $\Box$

   With respect to the usual topology defined by

$d((u_k)_{k\geq 1}, (v_k)_{k\geq 1}):= 2^{-min\{j\in \mathcal{N}:\,
u_j\neq v_j\}}$,\\ the pseudo-inverse operator is continuous because
it preserves  the prefix relation between two words. The following
result is of independent interest.
\begin{lem}\label{lem7}\rm
(1) The pseudo-inverse operator commutes with the limit operator,
that is, let $v\in \Sigma_n^{\omega}$, then
$v=\lim\limits_{k\rightarrow\infty} V_{k}$ and
\begin{eqnarray}
\Delta^{-1}_{\alpha}(\lim_{k\rightarrow\infty}
V_{k})=\lim_{k\rightarrow\infty} \Delta^{-1}_{\alpha}(V_{k}),\,
\text{where}\,\alpha\in \Sigma_n.\label{e4.4}
\end{eqnarray}

(2) Let $w=uv$, where $u\in \Sigma_n^{+},\;v\in \Sigma_n^{\omega}$,
then
\begin{equation}
\Phi^{-1}(w)=\Delta^{-|u|}_{u}(\Phi^{-1}(v)).\label{e4.5}
\end{equation}
\end{lem}
\prf{(1) Since $V_{k}=v[1\cdots k]$, we have
\begin{eqnarray}
\Delta^{-1}_{\alpha}(\lim_{k\rightarrow\infty}V_{k})
&=&\Delta^{-1}_{\alpha}(v[1\cdots k\cdots])\nonumber\\
&=&\alpha^{v_{1}}b_{i+1}^{v_{2}}b_{i+2}^{v_{3}}\cdots,\;\text{ where } \alpha=b_i\nonumber\\
&=&\lim_{k\rightarrow\infty} \Delta^{-1}_{\alpha}(V_{k}).\nonumber
\end{eqnarray}

(2) Let $u=u[1\cdots n]$  and $v=v[1\cdots k\cdots]$, then by
(\ref{e4.4}-\ref{e4.5}), we easily see that
\begin{eqnarray}
\Delta^{-|u|}_{u}(\Phi^{-1}(v))
&=&\Delta^{-|u|}_{u}(\lim_{k\rightarrow\infty}\Delta^{-(k-1)}_{v[1\cdots k-1]}(v[k]))\nonumber\\
&=&\lim_{k\rightarrow\infty}\Delta^{-(|u|+k-1)}_{uv[1\cdots k-1]}(v[k])\nonumber\\
&=&\lim_{k\rightarrow\infty}\Delta^{-(|u|+k-1)}_{w[1\cdots |u|+k-1]}(w[|u|+k])\nonumber\\
&=&\Phi^{-1}(w)\nonumber. \;\;  \Box
\end{eqnarray}}
Finally, it is easy to get
\begin{lem}\label{lem8}\rm
$w\in \Sigma^\omega$ is recurrent if and only if there are
infinitely many prefixes $\mathit{Pref}_{n_{i+1}}(w)$ of $w$ such
that $\mathit{Pref}_{n_{i+1}}(w)=x\cdot \mathit{Pref}_{n_i}(w)\cdot
y\cdot \mathit{Pref}_{n_i}(w)\cdot z$, where $x,\,y\in
\Sigma^*,\;z\in \Sigma^\omega,\; n_i<n_{i+1}$ for $i=1,2,3,\cdots$.
$\Box$
\end{lem}
\section{Recurrence of smooth infinite words\label{s7}}
Brlek et al. proved that smooth infinite words are recurrent for
2-letter alphabets having same parity (see Proposition 15 and
Theorem 25 in ~\cite{Br3}). We now establish that smooth infinite
words are recurrent for $n$-letter alphabets having the same
remainder when divided  by $n$ and the following proof seems to be
slightly more easily understood.
\begin{thm}\label{thm1}\rm
Let $\Sigma_n=\{a_1,a_2,\cdots,a_n\}$ with $a_i=n\cdot q_i+r,$ where
$0\leq r \leq n-1,\;q_i\in \mathcal{N}$ for $i=1,2,\cdots,n$. Then
every infinite smooth word over $\Sigma_n$ is recurrent.
\end{thm}
\prf{Since $\Phi$ is a bijection from $\mathcal{C}^{\omega}$ to
$\Sigma^{\omega}$, by  (\ref{e3.2}) and (\ref{e3.5}), for each
infinite smooth word $w$, there is a $u\in \Sigma^{\omega}$ such
that
\begin{eqnarray}
 w&=&\Phi^{-1}(u)\nonumber\\
  &=&\lim_{k\rightarrow \infty}\Delta_{u[1\cdots k-1]}^{-(k-1)}(u[k])\nonumber\\
  &=&\lim_{k\rightarrow \infty}\Phi^{-1}(U_k).\label{e5.1}
\end{eqnarray}
And note that $U_{k}=u[1\cdots k]$, from  (\ref{e3.3}) it
immediately follows that
\begin{eqnarray}
\Phi^{-1}(U_{k})&=&\Phi^{-1}(u[1\cdots k])\nonumber\\
                &=&\Delta^{-(k-1)}_{u[1\cdots k-1]}(u_{k})\nonumber\\
                &=&\Delta^{-(k-2)}_{u[1\cdots k-2]}((u^{n}_{k-1})^{\frac{u_{k}-r}{n}}u^{r}_{k-1})\nonumber\\
                &=&[\Delta^{-(k-2)}_{u[1\cdots k-2]}(u^{n}_{k-1})]^{\frac{u_{k}-r}{n}}\Delta^{-(k-2)}_{u[1\cdots
                k-2]}(u^{r}_{k-1}).
\end{eqnarray}

We next divide the proof into two cases in view of the remainder $r$
of $a_1,a_2,\cdots\text{ and } a_n$ when divided  by $n$.
And let $b_{k\cdot n+i}=b_i$ for $k\in \mathcal{N}$ and $i=1,2,\cdots,n$.\\
{\bf Case 1.} $r=0$. Then by Lemma \ref{lem5} (1), we have
\begin{eqnarray}
\Phi^{-1}(U_{k})&=&[\Delta^{-(k-2)}_{u[1\cdots
k-2]}(u^{n}_{k-1})]^{\frac{u_{k}}{n}}\nonumber\\
      &=&[\Delta^{-(k-3)}_{u[1\cdots k-3]}(\overbrace{u^{u_{k-1}}_{k-2}b^{u_{k-1}}_{i+1}\cdots b_{i-1+n}^{u_{k-1}}}^{n})]^{\frac{u_{k}}{n}},\text{ where } u_{k-2}=b_i\nonumber\\
      &=&[\Delta^{-(k-3)}_{u[1\cdots k-3]}(b^{u_{k-1}}_{i})\Delta^{-(k-3)}_{u[1\cdots k-3]}(b^{u_{k-1}}_{i+1})\cdots
      \Delta^{-(k-3)}_{u[1\cdots k-3]}(b^{u_{k-1}}_{i-1+n})]^{\frac{u_{k}}{n}}\nonumber\\
      &=&[\Delta^{-(k-4)}_{u[1\cdots k-4]}(\overbrace{u^{b_{i}}_{k-3}b^{b_{i}}_{j+1}\cdots
       b^{b_{i}}_{j-1+n}}^{u_{k-1}})\nonumber\\
       {}&& \Delta^{-(k-4)}_{u[1\cdots k-4]}(\overbrace{u^{b_{i+1}}_{k-3}b^{b_{i+1}}_{j+1}\cdots
       b^{b_{i+1}}_{j-1+n}}^{u_{k-1}})\cdots\nonumber
\end{eqnarray}
\begin{eqnarray}
\hspace{1.28cm}{}&&\Delta^{-(k-4)}_{u[1\cdots k-4]}(\overbrace{u^{b_{i-1+n}}_{k-3}b^{b_{i-1+n}}_{j+1}\cdots
       b^{b_{i-1+n}}_{j-1+n}}^{u_{k-1}})]^\frac{u_k}{n},\text{ where }u_{k-3}=b_j\nonumber\\
        &=&[\Delta^{-(k-4)}_{u[1\cdots k-4]}(u_{k-3})x_1\Delta^{-(k-4)}_{u[1\cdots k-4]}(u_{k-3})x_2\cdots \Delta^{-(k-4)}_{u[1\cdots k-4]}(u_{k-3})x_n]^{\frac{u_{k}}{n}}\nonumber\\
        &=&[\Phi^{-1}(U_{k-3})x_1\Phi^{-1}(U_{k-3})x_2\cdots \Phi^{-1}(U_{k-3})x_n]^{\frac{u_{k}}{n}},\nonumber
\end{eqnarray}
which suggests that $w$ is recurrent by Lemma~\ref{lem8}, where
$\Delta^{-(k-4)}_{u[1\cdots
k-4]}(\overbrace{u^{b_{i+l}}_{k-3}b^{b_{i+l}}_{j+1}\cdots
b^{b_{i+l}}_{j-1+n}}^{u_{k-1}})\\=\Delta^{-(k-4)}_{u[1\cdots
k-4]}(u_{k-3})x_l$ for $l=0,1,\cdots,n-1$ and $k=5,6,\cdots$.\\
{\bf Case 2.} $r>0$. Then, analogously, by Lemma \ref{lem5} (1), we
obtain
\begin{eqnarray}
\Phi^{-1}(U_{k})
          &=&[\Delta^{-(k-2)}_{U_{k-2}}(u_{k-1}^{n})]^{\frac{u_{k}-r}{n}}\Delta^{-(k-2)}_{U_{k-2}}(u_{k-1}^{r})\nonumber\\
          &=&[\Delta^{-(k-2)}_{U_{k-2}}(u_{k-1})x]^{\frac{u_{k}-r}{n}}\Delta^{-(k-2)}_{U_{k-2}}(u_{k-1})y\nonumber\\
          &=&[\Phi^{-1}(U_{k-1})x]^{\frac{u_{k}-r}{n}}\Phi^{-1}(U_{k-1})y,\label{e5.3}
\end{eqnarray}
where $\Delta^{-(k-2)}_{U_{k-2}}(u_{k-1}^{n})=\Delta^{-(k-2)}_{U_{k-2}}(u_{k-1})x$,
$\Delta^{-(k-2)}_{U_{k-2}}(u_{k-1}^{r})=\Delta^{-(k-2)}_{U_{k-2}}(u_{k-1})y$.\\
{\bf Case 2.1.} If there are infinitely many $u_k$ such that
$u_k>r$, then there exist infinitely many $u_k$ such that
$(u_k-r)/n\geq 1$. Thus from  (\ref{e5.3}) and Lemma~\ref{lem8}
it immediately follows that $w$ is recurrent.\\
{\bf Case 2.2.} $u=v\cdot r^\omega$ and $r=b_s$, where $v\in
\Sigma^*$. Then from Lemma~\ref{lem7} it follows that
\begin{eqnarray}
w=\Delta^{-|v|}_v(\Phi^{-1}(r^\omega)).\label{e5.4}
\end{eqnarray}
{\bf Case 2.2.1.} If $r\geq 2$, then by  (\ref{e5.4}) and
Lemma~\ref{lem7} (1), we get
\begin{eqnarray}
w&=&\Delta^{-|v|}_v(\lim_{k\rightarrow\infty}\Delta^{-(k-1)}_r(r))\nonumber\\
 &=&\lim_{k\rightarrow\infty}\Delta^{-|v|}_v(\Delta^{-(k-1)}_r(r))\nonumber\\
 &=&\lim_{k\rightarrow\infty}\Delta^{-|v|}_v(\Phi^{-1}(r^k)).\label{e5.5}
\end{eqnarray}
Let $q$ be the smallest positive integer such that the number of
runs of $\Delta_r^{-q}(r)$ is larger than $n$. Then
\begin{eqnarray}
\Delta^{-|v|}_v(\Phi^{-1}(r^k))&=&\Delta^{-|v|}_v(\Delta_r^{-(k-q-1)}(\Delta_r^{-q}(r)))\nonumber\\
 &=&\Delta^{-|v|}_v(\Delta_r^{-(k-q-1)}(r^{t_1}b_{s+1}^{t_2}\cdots b_{s+n-1}^{t_n}r^{t_{n+1}}\cdots))\nonumber\\
 &=&\Delta^{-|v|}_v(\Delta_r^{-(k-q-1)}(r^{t_1}b_{s+1}^{t_2}\cdots b_{s+n-1}^{t_n})\Delta_r^{-(k-q-1)}(r^{t_{n+1}}\cdots))\nonumber\\
 &=&\Delta^{-|v|}_v(\Delta_r^{-(k-q-1)}(r))x\Delta^{-|v|}_v(\Delta_r^{-(k-q-1)}(r))y\nonumber\\
 &=&\Delta^{-|v|}_v(\Phi^{-1}(r^{k-q}))x\Delta^{-|v|}_v(\Phi^{-1}(r^{k-q}))y,\label{e5.6}
\end{eqnarray}
where $t_i\in \Sigma_n$ and $\sum_{i=1}^nt_i$ is a multiple of $n$,

$\Delta^{-|v|}_v(\Delta_r^{-(k-q-1)}(r))x=\Delta^{-|v|}_v(\Delta_r^{-(k-q-1)}(r^{t_1}b_{s+1}^{t_2}\cdots
b_{s+n-1}^{t_n}))$,

$\Delta^{-|v|}_v(\Delta_r^{-(k-q-1)}(r^{t_{n+1}}\cdots))=\Delta^{-|v|}_v(\Delta_r^{-(k-q-1)}(r))y$.

Thus (\ref{e5.5}) and (\ref{e5.6}) mean that $w$ is recurrent by
Lemma~\ref{lem8}.

{\bf Case 2.2.2.} If $r=1$, then by  (\ref{e5.4}) and
Lemma~\ref{lem7} (1), we get
\begin{eqnarray}
w&=&\Delta^{-|v|}_v(\Phi^{-1}(1^\omega))\nonumber\\
 &=&\Delta^{-|v|}_v(1\Phi^{-1}(b_{s+1}^\omega))\nonumber\\
 &=&\Delta^{-|v|}_v(\lim_{k\rightarrow \infty}1\Phi^{-1}(b_{s+1}^k))\nonumber\\
 &=&\lim_{k\rightarrow \infty}\Delta^{-|v|}_v(1\Phi^{-1}(b_{s+1}^k)).\label{e5.7}
\end{eqnarray}
Thus
\begin{eqnarray}
\Delta^{-|v|}_v(1\Phi^{-1}(b_{s+1}^k))&=&\Delta^{-|v|}_v(1\Delta^{-(k-1)}_{b_{s+1}}(b_{s+1}))\nonumber\\
  &=&\Delta^{-|v|}_v(1\Delta^{-(k-2)}_{b_{s+1}}(b_{s+1}^{b_{s+1}}))\nonumber\\
  &=&\Delta^{-|v|}_v(1\Delta^{-(k-2)}_{b_{s+1}}(b_{s+1}^{b_{s+1}-1})\Delta^{-(k-2)}_{b_{s+1}}(b_{s+1})),\text{ by } n\mid (b_{s+1}-1)\nonumber\\
  &=&\Delta^{-|v|}_v((1\Delta^{-(k-2)}_{b_{s+1}}(b_{s+1}))x(1\Delta^{-(k-2)}_{b_{s+1}}(b_{s+1})))\nonumber\\
  &=&\Delta^{-|v|}_v(1\Phi^{-1}(b_{s+1}^{-(k-1)}))y\Delta^{-|v|}_v(1\Phi^{-1}(b_{s+1}^{-(k-1)})),\nonumber
\end{eqnarray}
which still implies that $w$ is recurrent by Lemma \ref{lem8} and
(\ref{e5.7}), where
$\Delta^{-(k-2)}_{b_{s+1}}(b_{s+1}^{b_{s+1}-1})\\=\Delta^{-(k-2)}_{b_{s+1}}(b_{s+1})x1$
and
$\Delta^{-|v|}_v(1\Delta^{-(k-2)}_{b_{s+1}}(b_{s+1})x)=\Delta^{-|v|}_v(1\Phi^{-1}(b_{s+1}^{-(k-1)}))y$.}
$\Box$
\section{Primitive substitution and uniform recurrence\label{s8}}
A substitution $\sigma$ is a function from the alphabet $\Sigma_n$
into the set $\Sigma_n^+$ of nonempty words; it can be extended to a
morphism of $\Sigma_n^*$ ($\Sigma_n^\omega$) in a natural way by
concatenation. A substitution $\sigma$ is primitive if there exists
a positive integer $k$ such that, for any $a$ and $b$ in $\Sigma_n$,
the letter $a$ occurs in $\sigma^k(b)$. About the periodic points of
a primitive substitution, we have the following simple fact.
\begin{prop}\rm(~\cite{Fogg} Proposition 1.2.3.)\label{prop1}
\; If $\sigma$ is primitive, then any of its periodic points is a
uniform recurrent sequence.  $\Box$
\end{prop}
Brelk et al.~\cite{Br3} showed that smooth infinite words are
recurrent for same parity alphabets, but we do not know wether or
not smooth infinite words are uniformly recurrent for same parity
alphabets. Note that $\Sigma_n$ is the alphabet with the cyclic
order $\mathbf{b_1b_2\cdots b_n}$. Now we consider the generalized
Kolakoski word $K_{c_1c_2\cdots c_n}$ over $\Sigma_n$ with the base
sequence $\mathcal{B}(K_{c_1c_2\cdots c_n})=(c_1c_2\cdots
c_n)^\omega$ and
\begin{eqnarray}\label{eq8.1}
c_i=q_i\cdot n+r\text{ for }i=1,2,\cdots,n.
\end{eqnarray}

For 2-letter alphabet $\Sigma_2=\{c_1,\;c_2\}$ having same parity,
if $c_1=2m$ and $c_2=2n$, where $m,n\in \mathcal{N}$, then
$K_{c_1c_2}$ is the fixed point of the following primitive
substitution which was given by Sing \cite{Sing2}:
\begin{eqnarray}\label{e9.5}
\sigma:\;\begin{array}{ll}
A\mapsto  A^mB^m\\
B\mapsto  A^nB^n,\\
\end{array}\nonumber
\end{eqnarray}
where $A=c^2_1$ and $B=c^2_2$.

If $c_1=2m+1$ and $c_2=2n+1$, where $m<n\in \mathcal{N}$, then
$K_{c_1c_2}$ is the fixed point of the following primitive
substitution $\mu$ which Sing constructed in \cite{Sing3}:
\begin{eqnarray}\label{e9.5}
\mu:\;\begin{array}{ll}
A\mapsto  A^mBC^m\\
B\mapsto  A^mBC^n\\
C\mapsto  A^nBC^n,\\
\end{array}\nonumber
\end{eqnarray}
where $A=c^2_1$, $B=c_1c_2$ and $C=c^2_2$.

Thus Proposition \ref{prop1} gives the generalized kolakoski
sequences are uniformly recurrent for 2-letter alphabets having same
parity.

 In this section, when $r=0$, or $r>0$ and $n$ is an even
number, we construct a primitive substitution $\sigma$ of $\Sigma_n$
such that $K_{c_1c_2\cdots c_n}$ is the fixpoint of $\sigma$.\\
{\bf Case 1.} $r=0$ in (\ref{eq8.1}).

In this case, the letters $c_1,\,c_2,\,\cdots,\,c_n$ are all the
multiples of $n$, then set $A_i=c_i^n$ for $i=1,2,\cdots,n$, and
the substitution $\sigma$ is given by
\begin{eqnarray}
\sigma:\;\begin{array}{ll} A_i\mapsto A_1^{q_i} A_2^{q_i}\cdots
A_n^{q_i} \text{ for } i=1,2,\cdots,n.
\end{array}\label{eq8.2}
\end{eqnarray}
Note that $q_i>0$, from (\ref{eq8.2}) it immediately follows that
$\sigma$ is both a primitive substitution and  $K_{c_1c_2\cdots
c_n}=\lim_{t\rightarrow \infty}\sigma^t(A_1)$, which means that
$K_{c_1c_2\cdots c_n}$ is a fixpoint of the primitive substitution
(\ref{eq8.2}).\\
{\bf Case 2.} $r>0$ in (\ref{eq8.1}) and $n$ is
an even number with $n=2m$, where $m\in \mathcal{N}$.

Set $A_i=c_i^n$ for $i=1,2,\cdots,n$, $B_i=c_{2i-1}^rc_{2i}^r$ for
$i=1,2,\cdots,m$.\\
{\bf Case 2.1.} $r=2h,\,h\in \mathcal{N}$. Then the corresponding
substitution $\sigma$ are determined by
\begin{eqnarray}\label{eq8.3}
\sigma:\;\begin{array}{lcl}
A_{2k+1}\mapsto  A_{2k\cdot r+1}^{q_{2k+1}}B_{k\cdot r+1}A_{2k\cdot r+2}^{q_{2k+1}}A_{2k\cdot r+3}^{q_{2k+1}}B_{k\cdot r+2}A_{2k\cdot r+4}^{q_{2k+1}}\cdots A_{2k\cdot r+n-1}^{q_{2k+1}}B_{k\cdot r+m}A_{2k\cdot r+n}^{q_{2k+1}}\\
B_{k+1}\mapsto  A_{2k\cdot r+1}^{q_{2k+1}}B_{k\cdot r+1}A_{2k\cdot r+2}^{q_{2k+1}}\cdots A_{2(k\cdot r+h)-1}^{q_{2k+1}}B_{k\cdot r+h}A_{2(k\cdot r+h)}^{q_{2k+1}}A_{2(k\cdot r+h)+1}^{q_{2(k+1)}}\\
\hspace{1.69cm} B_{k\cdot r+h+1}A_{2(k\cdot r+h+1)}^{q_{2(k+1)}}\cdots  A_{2(k+1)r-1}^{q_{2(k+1)}}B_{(k+1) r}A_{2(k+1)r}^{q_{2(k+1)}}\\
A_{2(k+1)}\mapsto A_{2(k+1)r+1}^{q_{2(k+1)}}B_{(k+1)r+1}A_{2(k+1)r+2}^{q_{2(k+1)}}\cdots  A_{2(k+1)r+n-1}^{q_{2(k+1)}}B_{(k+1)r}A_{2(k+1)r+n}^{q_{2(k+1)}}\;,\\
\end{array}
\end{eqnarray}
where $k=0,1,2,\cdots,m-1$, $A_i=A_j$ if $i\equiv j\,(\text{mod}\;
n)$, $B_i=B_j$ if $i\equiv j\,(\text{mod}\; m)$.

For example, if $\Sigma=\{2,6,10,14\}$ then $n=4,r=2$. Thus
$A_1=6^4$, $A_2=10^4$, $A_3=14^4$, $A_4=2^4$, $B_1=6^210^2$,
$B_2=14^22^2$. From (\ref{eq8.3}) it follows that
\begin{eqnarray}\label{e9.5}
\sigma_1:\;\begin{array}{ll}
A_{1}\mapsto  A_1B_1A_2A_3B_2A_4\\
B_{1}\mapsto  A_1B_1A_2A_3^{2}B_2A_4^2\\
A_{2}\mapsto  A_1^2B_1A_2^2A_3^2B_2A_4^2\\
A_{3}\mapsto  A_1^3B_1A_2^3A_3^3B_2A_4^3\\
B_{2}\mapsto  A_1^3B_1A_2^3B_2\\
A_4\mapsto  B_1B_2\\
\end{array}.
\end{eqnarray}
By (\ref{e9.5}), we have
\begin{eqnarray}\label{eq1}
A_1&\mapsto& 6^610^614^62^6\mapsto
6^610^614^62^66^610^614^{10}2^{10}6^{10}10^{10}14^{10}2^{10}\nonumber\\
&&6^{14}10^{14}14^{14}2^{14}6^{14}10^{14}14^22^26^210^214^22^2\mapsto\cdots\nonumber\\
&=&6^610^614^62^66^610^614^{10}2^{10}6^{10}10^{10}14^{10}2^{10}\nonumber\\
&&6^{14}10^{14}14^{14}2^{14}6^{14}10^{14}14^22^26^210^214^22^2\cdots=K_{6(10)(14)2}.
\end{eqnarray}

From (\ref{e9.5}) it easily follows that for any $u$ and $v$ in
the alphabet $\{2,6,10,14\}$, $u$ must occur in $\sigma^3(v)$,
which suggests that $\sigma$ is a primitive substitution over the
alphabet $\{2,6,10,14\}$. Moreover, from (\ref{eq1}) we have
$K_{6(10)(14)2}=\lim_{t\rightarrow \infty}\sigma_1^t(A_1)$. Thus
$K_{6(10)(14)2}$ is a fixpoint of the primitive substitution
(\ref{e9.5}).

In general, note that at most one of $q_1,q_2,\cdots\text{ and
}q_n$ can take the value 0, from (\ref{eq8.3}) we easily see that
for any $u$ and $v$ in the alphabet
$\{A_1,A_2,\dots,A_n,B_1,B_2,\cdots,B_m\}$, $u$ must occur in
$\sigma^3(v)$, which suggests that $\sigma$ is a primitive
substitution and $K_{c_1c_2\cdots c_n}=\lim_{t\rightarrow
\infty}\sigma^t(A_1)$. Therefore $K_{c_1c_2\cdots c_n}$ is  a
fixpoint of the primitive substitution (\ref{eq8.3}).\\
{\bf Case 2.2.} $r=2h+1,\,h\in N$. Then the corresponding
substitution $\sigma$ are determined by
\begin{eqnarray}\label{e8.3}
\sigma:\;\begin{array}{ll}
A_{2k+1}\mapsto  A_{2k\cdot r+1}^{q_{2k+1}}B_{k\cdot r+1}A_{2k\cdot r+2}^{q_{2k+1}}A_{2k\cdot r+3}^{q_{2k+1}}B_{k\cdot r+2}A_{2k\cdot r+4}^{q_{2k+1}}\cdots A_{2k\cdot r+n-1}^{q_{2k+1}}B_{k\cdot r+m}A_{2k\cdot r+n}^{q_{2k+1}}\\
B_{k+1}\mapsto  A_{2k\cdot r+1}^{q_{2k+1}}B_{k\cdot r+1}A_{2k\cdot r+2}^{q_{2k+1}}\cdots A_{2(k\cdot r+h)-1}^{q_{2k+1}}B_{k\cdot r+h}A_{2(k\cdot r+h)}^{q_{2k+1}}A_{2(k\cdot r+h)+1}^{q_{2k+1}}\\
\hspace{1.69cm} B_{k\cdot r+h+1}A_{2(k\cdot r+h+1)}^{q_{2(k+1)}}\cdots  A_{2(k+1)r-1}^{q_{2(k+1)}}B_{(k+1) r}A_{2(k+1)r}^{q_{2(k+1)}}\\
A_{2(k+1)}\mapsto A_{2(k+1)r+1}^{q_{2(k+1)}}B_{(k+1)r+1}A_{2(k+1)r+2}^{q_{2(k+1)}}\cdots  A_{2(k+1)r+n-1}^{q_{2(k+1)}}B_{(k+1)r}A_{2(k+1)r+n}^{q_{2(k+1)}}\;,\\
\end{array}
\end{eqnarray}
where $k=0,1,2,\cdots,m-1$, $A_i=A_j$ if $i\equiv j\,(\text{mod}\;
n)$, $B_i=B_j$ if $i\equiv j\,(\text{mod}\; m)$.

For example, if $\Sigma=\{1,5,9,13\}$ then $n=4,r=1$. Thus
$A_1=5^4$, $A_2=9^4$, $A_3=13^4$, $A_4=1^4$, $B_1=59$, $B_2=(13)1$.
From (\ref{e8.3}) it follows that
\begin{eqnarray}\label{e9.4}
\sigma_2:\;\begin{array}{ll}
A_{1}\mapsto  A_1B_1A_2A_3B_2A_4\\
B_{1}\mapsto  A_1B_1A_2^2\\
A_{2}\mapsto  A_3^2B_2A_4^2A_1^2B_1A_2^2\\
A_{3}\mapsto  A_3^3B_2A_4^3A_1^3B_1A_2^3\\
B_{2}\mapsto  A_3^3B_2\\
A_{4}\mapsto  B_1B_2\\
\end{array}.
\end{eqnarray}
In view of (\ref{e9.4}), we get
\begin{eqnarray}\label{eq2}
A_1&\mapsto& 5^59^513^51^5\mapsto
5^59^513^51^55^59^913^91^95^99^913^{13}1^{13}5^{13}9^{13}13^{13}159(13)1\mapsto\cdots\nonumber\\
&=&5^59^513^51^55^59^913^91^95^99^913^{13}1^{13}5^{13}9^{13}13^{13}159(13)1\cdots=K_{59(13)1}.\nonumber
\end{eqnarray}

By an argument similar to Case 2.1, from (\ref{e8.3}) we obtain
that $K_{c_1c_2\cdots c_n}$ is  a  fixpoint of the primitive
substitution (\ref{e8.3}).

Up to now, we have proved that if  $r=0$ or $r>0$ and $n$ is an even
number, then $K_{c_1c_2\cdots c_n}$ is a fixpoint of the primitive
substitution $\sigma$.  Thus from Proposition~\ref{prop1} it
immediately follows that $K_{c_1c_2\cdots c_n}$ is uniformly
recurrent, which suggests that
\begin{thm}\rm\label{thmm}
Let $n$ be an integer larger than 1, and $a_i=q_i\cdot n+r$ for
$i=1,2,\cdots,n,\;0\leq r \leq n-1$. If $r=0$ or $r>0$ and $n$ is an
even number, then the generalized Kolakoski words over the
$n$-letter alphabet $\Sigma_n$ with the cyclic order
$\mathbf{b_1b_2\cdots b_n}$ are the uniform recurrent sequences.
$\Box$
\end{thm}

\section{Reversal\label{s9}}
Brlek et al. proved that the set $F(w)$ is closed under reversal for
2-letter odd alphabets by ~\cite{Br3} Lemma 11. For $n$-letter
alphabet $\Sigma_n$ with the cyclic order $\mathbf{b_1b_2\cdots
b_n}$, if $n>2$, then the reversal of any factor of
$\mathbf{b_1b_2\cdots b_n}$ of length larger than 1 is not
compatible with $\mathbf{b_1b_2\cdots b_n}$. Thus the reversal of
any factor of a smooth infinite word $w$ with the number of runs
being larger than 1 must be not again a factor of $w$. Thus closure
property of reversal of factors of smooth infinite words is of
meaningful only for 2-letter alphabets. Since the proof of
~\cite{Br3} Lemma 11 seems to be somewhat complicated, we now give a
different proof, which is slightly more explicit.

\begin{thm}\label{thm4}\rm(\cite{Br3} Proposition 15)\;
Let $w$ be a smooth infinite word over the alphabet $\Sigma_2$. If
$a_1,\;a_2$ are odd integers, then the set $F(w)$ is closed under
reversal.
\end{thm}
\prf{Since $w=\Phi^{-1}(u)$ for some $u\in \Sigma_n^{\omega}$, by
 (\ref{e3.4}), $w=\lim\limits_{k\rightarrow
\infty}\Phi^{-1}(u[1\cdots k])$. Note that $u[k]$ is a palindrome of
length 1 and $a,\;b$ are odd integers. So by  (\ref{e3.3}) and Lemma
\ref{lem3} (3), we see that $\Phi^{-1}(u[1\cdots
k])=\Delta_{u[1]}^{-1}\Delta_{u[2]}^{-1}\cdots\Delta_{u[k-1]}^{-1}(u[k])$
is a palindrome of odd length. Since for every $f\in F(w)$, there is
a positive integer $m$ such that $\Phi^{-1}(u[1\cdots m])=gfq$, so
$\Phi^{-1}(u[1\cdots m])=\widetilde{\Phi^{-1}(u[1\cdots
m])}=\tilde{q}\tilde{f}\tilde{g}$, that is, $\tilde{f}\in F(w)$,
which guarantees that $F(w)$ is closed under reversal.}  $\Box$
\section{Concluding remarks\label{s11}}
Brelk et al.~\cite{Br3} showed that smooth infinite words are
recurrent for 2-letter alphabets having same parity. In the
section~\ref{s7}, we show that smooth infinite words are recurrent
for $n$-letter alphabets $\Sigma_n$ having same remainder when
divided by $n$. In the section~\ref{s8}, we establish that the
generalized Kolakoski words over the $n$-letter alphabet $\Sigma_n$
are uniformly recurrent except for the case $r>0$ and $n$ being an
odd integer.

Thus for the alphabets $\Sigma_n$ having same nonzero remainder when
divided by $n$ and $n$ being a positive odd integer, to determine
whether or not the generalized Kolakoski words  are uniformly
recurrent is a fascinating problem. In general,  to ascertain
whether smooth infinite words over $n$-letter alphabet $\Sigma_n$
are uniformly recurrent also deserves further investigations.
Similarly, letter frequency and permutation invariant property of
smooth infinite words  also merits further explorations.

Moreover, by corollary~\ref{c2}, if $n\mid a_i$ for
$i=1,2,\cdots,n$, then the letter frequency  of generalized
Kolakoski words  is $1/n$ for the alphabet $\Sigma_n$ with given
cyclic order.

In addition, let $n=2m$, $a_i=n\cdot q_i+r$ for $i=1,2,\cdots,n$ and
$r=m$. And if we could construct a primitive substitution $\sigma$
of constant length $n$ over the alphabet
$\{A_1,A_2,\dots,A_n,B_1,B_2,\cdots,B_m\}$, then
from~\cite{Queffelec} Proposition V.9. it follows that each letter
in the fixed point $u=\lim_{i\rightarrow \infty}\sigma^i(A_1)$ of
$\sigma$, occurs in $u$ with a positive frequency. Let $p_i$ be the
frequency of $A_i$ occurring in $u$ for $i=1,2,\cdots,n$ and $q_i$
be the frequency of $B_i$ occurring in $u$ for $i=1,2,\cdots,m$,
then by $A_i=c_i^n$ and $B_j=c_{2j-1}^mc_{2j}^m$, we see that the
letter $c_{2i-1}$ occurs exactly in $A_{2i-1}$ and $B_i$, and
$c_{2i}$ occurs exactly in $A_{2i}$ and $B_i$. Thus the frequency of
$c_{2i-1}$ occurring in $K_{c_1c_2\cdots c_n}$ is equal to
$p_{2i-1}+1/2\cdot q_i$, and the frequency of $c_{2i}$ occurring in
$K_{c_1c_2\cdots c_n}$ is equal to $p_{2i}+1/2\cdot q_i$ for
$i=1,2,\cdots,m$.  Therefore we could arrive at the following
attractive result:

For the $n$-letter alphabet $\Sigma_n$ with the cyclic order, if
$r>0,n=2m$ and $r=m$, the letters occurred in the generalized
Kolakoski words have the positive frequency.

Thus the following open problem is very significative.

{\bf Prove (or disprove)} that for the generalized Kolakoski
sequence $K_u$ over $n$-letter alphabet $\Sigma_n$, there exists a
primitive substitution $\sigma$ of length constant over the alphabet
$\{A_1,A_2,\dots,A_n,B_1,B_2,\cdots,B_m\}$ such that $K_u$ is a
fixed point of $\sigma$.

\flushleft 
\end{document}